# Structure and thermodynamics of associating rods solutions

Min Sun Yeom, Alexander V. Ermoshkin and Monica Olvera de la Cruz*
*Department of Materials Science and Engineering, Northwestern University,
Evanston, Illinois 60208-3108*
*\* Corresponding author, m-olvera@northwestern.edu*

PACS. 61.30.Cz   Molecular and microscopic models and theories of liquid crystal structure

**Abstract**. - Thermoreversible sol-gel transitions in solutions of rod-like associating polymers are analyzed by computer simulations and by mean field models. The sol-gel transition is determined by the divergence of the cluster weight average. The analytically determined sol-gel transition is in good agreement with the simulation results. At low temperatures we observe a peak in the heat capacity, which maximum is associated with the precipitation transition. The gelation transition is sensitive to the number of associating groups per rod but nearly insensitive to the spatial distribution of associating groups around the rod. The precipitation is strongly dependent on both the number and distribution of associating groups per rod. We find negligible nematic orientational order at the gelation and precipitation transitions.

   The synthesis and design of rod-like molecules with strong specific interactions along their backbones has opened the possibility of creating new materials with unique physical properties. Examples of such systems include peptide amphiphilic molecules self-organized into nanofibers which can be reversibly crosslinked [1] and genetically engineered helixes with specific distributions of reactive groups along their backbones [2]. The structure of rod like molecules with designed strong interactions along their backbones may be quite complex. Here we analyze the structure and thermodynamics of rod-like molecules as a function of the number and distribution of associating groups along their backbones.
   Monodispersed solutions of rodlike molecules without attractive interactions have been extensively studied analytically [3-6] and by computer simulation [7-13]. These systems can form various thermodynamic phases including isotropic, nematic, smetic and solid phases. The transitions between these phases depend on the concentration and aspect ratio of the rod like molecules, and they are modified in the presence of attractive interactions [14]. Strong physical interactions at specific sites along the rod backbone such as hydrogen bonding, is expected not only to modify the transitions lines, but also to open the possibility of creating new phases.  In particular, one may expect the formation of gel-like structures.
   Thermoreversible gelation of flexible polymers has recently received a lot of attention [15-30]. Theoretical [15-25] and experimental [26-30] studies have concentrated at determining the formation of an infinite network of macromolecules which is cross-linked via physical interactive sites. It is difficult to theoretically predict the sol-gel transition of associating rigid polymer solutions because these systems have strong correlations that give the unique possibility of phase separation between isotropic gels and nematic-like gels or gels of bundles of chains.  Experimental evidence suggests that there is phase separation when physically associating rigid polymer solutions are cooled [26].
   In this paper we study the phase behavior of rod polymer solutions with physical interaction at specific sites along their backbones by Monte Carlo simulations and compare the results to mean field models. We assume constant number of particles (N), volume (V) and temperature (T).

Periodic boundary conditions, using the minimum image convention, are applied in all directions, and the conventional Metropolis algorithm is used. We consider hard rods that we construct as connected hard spheres, which are referred to as monomers, with various numbers of stickers per rod. The stickers can only associate with each other. The stickers diameter ($\sigma_s$) is a half of that of the monomers. We model four types of rods with different number and/or distribution of stickers per rod, labeled I to IV in Fig. 1. Two stickers were put on the 2$^{nd}$ and 9$^{th}$ monomers on the same side (Type I) and on opposite sides (Type II) of the rod. Five stickers were put on the odd number monomers along the same direction (Type III) and evenly distributed along five different orientations rotated by 72 degrees (Type IV). The rightmost picture in Fig. 1 is the top view of type IV rod. The stickers have hard core repulsions and square well attractive interactions ($-\varepsilon$) in the close vicinity of each other (0.05 $\sigma_s$). The reduced density and temperature are defined as $\rho^* = 10 N_m \sigma^3 / V$, $T^* = k_B T / \varepsilon$, where $N_m$ is the number of molecules (rods) and $\sigma$ is the diameter of the monomers. The stickers are excluded when $\rho^*$, dimensionless mean energy per monomer, and other properties are calculated.

In order to locate the gelation line and the phase transition lines, computer simulations were performed at various reduced densities at $T^*$ between 0.1 and 0.8. Two molecules are called "connected" if there is at least one sticker pair within the cut off distance (1.05 $\sigma_s$). We define a set of connected molecules as a cluster. At the gelation (or percolation) transition the cluster weight average $M_w$ diverges. Fig. 2 shows the gelation lines and precipitation transition lines for the four types of molecules, where the precipitation lines are determined by the maximum in the peak of the dimensionless heat capacity $C^*$ as a function of $T^*$.

First we discuss the gelation lines. Near the percolation threshold, $M_w$ obeys the following scaling law, [31]

$$M_w = \sum_s s^2 n_s \Big/ \sum_s s n_s \sim |p - p_c|^{2\tau - 6} \tag{1}$$

Here $n_s$ is the average number of clusters of size $s$. The geometrical exponent $\tau$ is universal, independent of the microscopic details of the system. In Fig. 3 we show the variation of $M_w$ versus $T^*$ for type IV at $\rho^* = 0.3$. The straight line has a slope which corresponds to an effective exponent $\tau \sim 2.6$. The percolation exponents in three dimensions and in the Bethe lattice are 2.18 and 2.5, respectively [31]. Although the exponent in our finite size systems largely fluctuates for the different densities, the geometrical exponents have similar values for all types of rods, as expected.

We use mean filed theory to calculate the gelation points analytically. We assume that all finite size clusters in the system have only tree-like configurations. We also propose that the multiplicities of the junctions formed by stickers can not be greater then 3 for molecules of type I and II (see Figure 2a) and it can only be 2 for type III and IV. Since the number of stickers per chains $f$ is equal to 2 in rods type I and II, only permitting multiplicity 2 in the mean field model will lead to a linear chain of infinite degree of polymerization at the gelation point. In this case the sol-gel transition occurs at infinite density or at $T^*=0$ (see Eq. (14) with $f=2$ and $w_2$ given in Eq. (11)), which disagrees with the computational results. Therefore, triple sticker contacts have to be included in these types of molecules. On the other hand, as discussed below, for molecules type III and IV, the gelation is reasonably well described with only binary sticker contacts given that the number of 3 stickers in contact is very reduced at low densities.

For solutions of molecules type I and II we find the densities $\rho_1$, $\rho_2$ and $\rho_3$ of free stickers, 2-fold and 3-fold aggregates, respectively, by applying the law of mass action (valid for tree-like architectures) in the form

$$\frac{\rho_2}{\rho_1^2} = \frac{w_2}{2!}, \quad \frac{\rho_3}{\rho_1^3} = \frac{w_3}{3!} \tag{2}$$

where $w_2$ ($w_3$) is the statistical weight of 2(3)-fold junction with 2!(3!) symmetry index. Introducing the fractions

$$\Gamma_i = i\rho_i/\rho \ (i = 1,2,3) \tag{3}$$

where $\rho$ is the total density of stickers in the system, we rewrite Eq. (2) in the form

$$\Gamma_2/\Gamma_1^2 = w_2\rho, \ \Gamma_3/\Gamma_1^3 = \rho^2 w_3/2, \ \Gamma_1 + \Gamma_2 + \Gamma_3 = 1 \tag{4}$$

We identify the gelation point as the formation of an infinitely large network characterized by the divergence of the cluster weight average $N_w \to \infty$. Applying the diagram technique described in Ref. [21] we assign fugacity $z$ to each molecule and write the total concentration of molecules $\rho_{mol} = \rho/2$ in the form

$$\rho_{mol} = zt^2/2 \tag{5}$$

Here $t = t(z)$ is the concentration of all different branches (including the case of an empty branch) that can be attached to one sticker. The function $t(z)$ satisfies the following recurrent relation

$$t = 1 + w_2 zt + w_3(zt)^2/2 \tag{6}$$

The densities $\rho_1$, $\rho_2$ and $\rho_3$ of free stickers, 2-fold and 3-fold aggregates respectively can be written as

$$\rho_1 = zt, \ \rho_2 = w_2(zt)^2/2!, \ \rho_3 = w_2(zt)^3/3! \tag{7}$$

It can be shown explicitly that the condition $N_w \to \infty$ for the gelation point can be replaced with

$$\frac{\partial \rho_{mol}}{\partial z} \to \infty \tag{8}$$

Using Eqs. (5) and (6) we obtain that Eq. (8) is satisfied only if

$$1 - w_2 z - w_3 z^2 t = 0 \tag{9}$$

By taking into account the definitions of $\Gamma_i$ ($i = 1,2,3$) given by Eqs. (3), we obtain the following expression that defines the gelation point in the system

$$\Gamma_1 + 2\Gamma_2 + 3\Gamma_3 = 2 \tag{10}$$

In order to find the gelation temperature as a function of the density of molecules $\rho_{mol}$ we write $w_2(T)$ and $w_3(T)$ in the following form

$$w_2 = v_b e^{\varepsilon/k_B T}, \ w_3 = w_2^2 \tag{11}$$

where $v_b$ and $\varepsilon$ is the volume and energy of an associated pair respectively. The volume $v_b$ is approximately $\pi/12$ which is one half of the excluded volume of a single sticker.

By numerically solving Eqs (4), (10), (11) we obtain the curve shown in Fig. 2a. Though the analytic results are slightly underestimated, they follow the same trends as the simulation results at dilute solutions. The curves can be move up to match quantitatively better the simulation results if we modify $w_3(T)$ in Eq. (11) to account for a different effective volume for three associated stickers.

The gelation points for solutions of molecules type III and IV are obtained by using the mean field theory for pair wise association [19,20]. The law of mass action can be written in the form

$$\Gamma_2/\Gamma_1^2 = w_2\rho, \ \Gamma_1 + \Gamma_2 = 1 \tag{12}$$

The fraction $\Gamma_2$ of associated stickers at the gelation point is

$$\Gamma_2 = 1/{f-1} \tag{13}$$

where the number of stickers per molecule $f = 5$. Substituting Eq. (6) into Eq. (5) we obtain an analytical relation between the temperature and density of molecules at the gelation point

$$\rho_{mol} = \frac{1}{w_2(T)} \frac{f-1}{f(f-2)^2} \tag{14}$$

where $w_2(T)$ is defined by Eq. (11). Figure 2b shows that the simplest mean field theory used to describe type III and IV molecules is in reasonable agreement with the simulation results. The theory slightly overestimates the simulation results at low densities and underestimates the results at intermediate densities. This simple mean field theory is insensitive to the distribution of stickers per molecule and it ignores excluded volume effects. Even though we neglect the possibility of three or more associated stickers forming pairs, the fraction of pairs that are formed with more than two stickers is small at dilute solutions. This is evident by the lack of nematic order at the gelation transition. At intermediate densities, however, molecules type III show more degree of local orientational order, explaining the larger deviations from the simplest gelation theory with only binary contacts.

The precipitation lines shown in Fig. 2 are always located under the gelation line. In Fig. 4a we show the dimensionless heat capacity $C^*$ as a function of $T^*$ for molecules type III at various densities. For $\rho^* = 0.0005$ the peak is expected to occur at $T^* < 0.1$. In infinite systems first order transitions are characterized by delta function singularities in the second derivatives of thermodynamic potentials, such as specific heat. In finite systems, however, delta function singularities are rounded off. The maximum at the rounded peak of the specific heat represents the smeared delta function.[32] Notice that in Fig. 4a the peak in narrows as we diluted the system, which in constant number of particle simulations this is equivalent to increase the system size. This suggests that in an infinite system the peak in $C^*$ will go to a delta function. Therefore, the precipitation transitions shown in Fig. 2, which are determined by the maximum in the curve of $C^*$ versus $T^*$, are first order transition. It is argued that gelation is not a thermodynamic transition [20]. Recently, however, it has been proposed that the sol-gel transition is a genuine first-order phase transition due to mesoscopic cyclization effects that take place in the gel phase [22]. Here we do not find thermodynamic singularities in associating rod solutions at the sol-gel transition points.

In Fig. 4b we show $C^*$ as a function of $\rho^*$ at $T^* = 0.2$. For molecules types I and II there is no precipitation transition at $T^* = 0.2$. As shown in Fig. 2a, we only find precipitation transitions at high reduced densities ($\rho^* = 0.4$) for these molecules for the range of $T^*$ that we access here ($T^* > 0.1$). For type III and IV molecules we find a peak which occurs at much smaller reduce densities for molecules type III than for molecules type IV. These differences are explained below by correlating the precipitation transition to the degree of ordering in the clusters.

We calculate the second order Legendre polynomial $P_2$ around the gelation and phase transition lines to quantify the orientational order of the system and clusters,

$$P_2(\cos\theta_{ij}(r)) = (3\cos\theta_{ij}(r) - 1)/2 \qquad (15)$$

There is no overall nematic order for all molecule types solutions at $\rho^*$ between 0.0005 and 0.4 and $T^*$ between 0.1 and 0.8. The second Legendre polynomials for the clusters, however, show their structures depend on the types of molecules, densities and temperatures. The type III molecules orient locally at specific temperatures and densities. Types II and IV molecules form isotropic networks. Although types I and III may prefer forming long alignments, the limited simulation box size hinder the clusters into forming long alignments. This is why these molecules types have smaller size clusters than their corresponding types II and IV, respectively. The energy of the system decreases when the second Legendre polynomials for the clusters increase. The local or cluster orientational order $P_2$ and the $C^*$ at low $T^*$ (below the gelation line) have the same trends as a function of $\rho^*$, shown in Fig. 4b for the heat capacity. The origin of the maximum in either quantity $P_2$ or heat capacity can be traced to a competition between entropic and energetic effects at low temperatures and the change of the cluster structure from isotropic networks to clusters of oriented rods (bundles) as the density increases. The region where clusters of oriented rods are observed is not an equilibrium phase. At the precipitation transition equilibrium is established between two coexisting phases (a miscibility gap) of different reduced densities; the dilute phase is a network phase and the concentrated phase is a phase with an overall orientational order. Since the precipitations transitions occur at low $T^*$ for type IV molecules a crystalline phase may form in these molecules. With our simulations, however, we only access the rod reduced density network phase. The equilibrium reduced density of the nematic (and/or solid) phase is expected at $\rho^*$ values higher than the ones studied here. Various initial conformations including isotropic and phases with different degrees of nematic orientational order were used to test our results.

In summary, the gelation line in dilute solutions of rod-like associating polymers is rather insensitive to the distribution of associating groups along the rods. The gelation is strongly dependent on the number of associating groups per rod. We find no thermodynamic singularities at the gelation line. The gel points obtained by the mean field theory of gelation are in good agreement with our simulation results. The precipitation transitions are strongly dependent on the number and distribution of stickers per rod. We find a strong correlation between the degree of orientational order in the clusters and the precipitation transition.

∗∗∗

We acknowledge the financial support of the NIH grant number GM62109-02. MO and AE acknowledge partial financial support by the NSF grants DMR-0109610 and EEC-0118025, and by the Air Force Research Laboratory, Materials and Manufacturing Directorate, Wright-Patterson Air force Base.

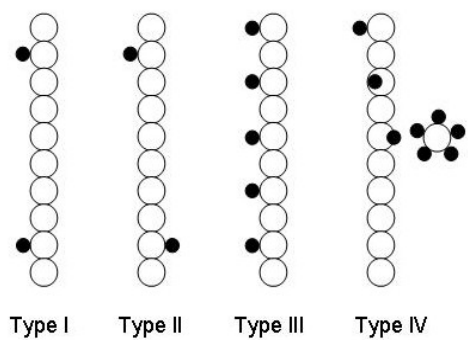

Fig. 1  The four types of rods showing the number and distribution of stickers per rod.

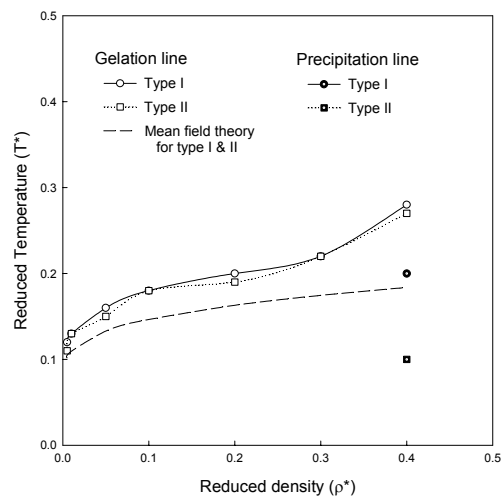

Fig. 2a  Gelation and phase transition lines for solution of rods with 2 stickers.

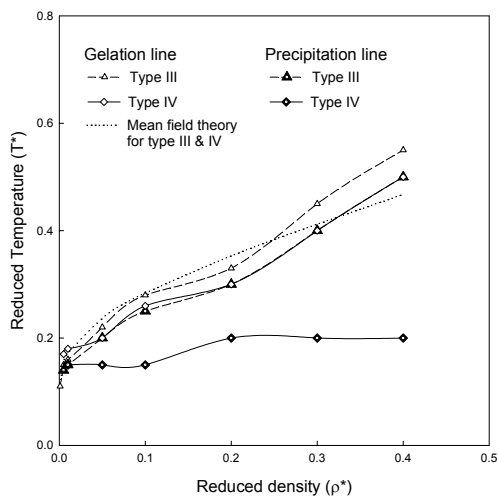 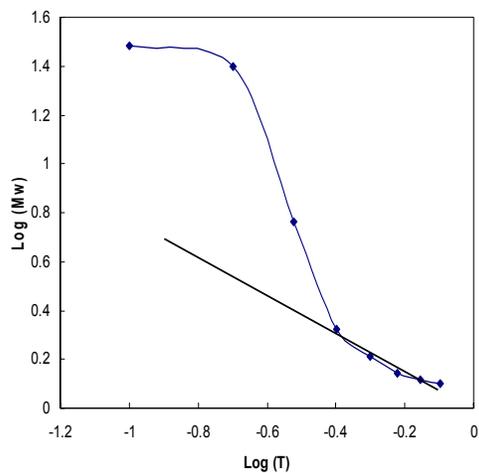

Fig. 2b Gelation and phase transition lines for solution of rods with 5 stickers. (The simulation error is obtained from the standard deviation estimate from three independent simulation runs for the internal energy; for molecules type III, for example, the error is 5% at $\rho^* = 0.005$ and $T^* = 0.15$).

Fig. 3 Variation of the mean cluster size versus $T^*$ for type IV molecules at $\rho^* = 0.3$.

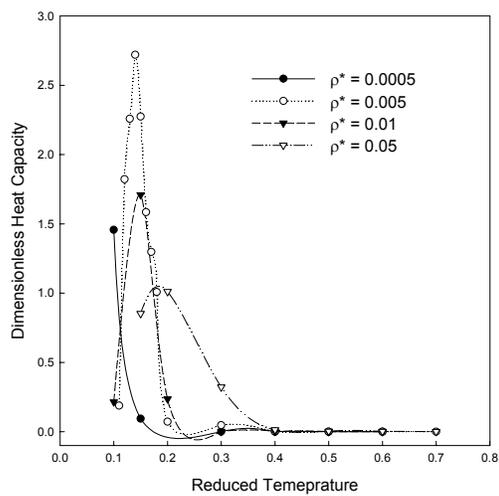 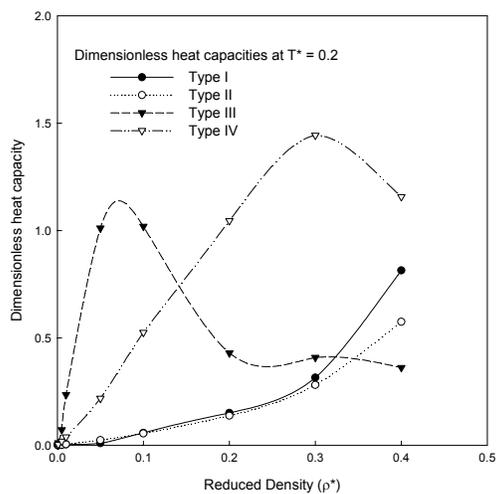

Fig. 4a Dimensionless heat capacity as a function of $T^*$ for type III molecules at various densities.

Fig. 4b Dimensionless heat capacity as a function of $\rho^*$ at $T^* = 0.2$.